\documentclass[onecolumn]{svjour3}

\RequirePackage{fix-cm}
\usepackage{amssymb}
\usepackage[T1]{fontenc}
\usepackage{textcomp}
\usepackage{amstext}
\usepackage{graphicx}
\usepackage[biblabel]{cite}
\usepackage{epsfig}
\usepackage{amsfonts}
\usepackage{amsmath}
\usepackage[english]{babel}
\usepackage{subfigure}
\usepackage{color}
\graphicspath{./}
\usepackage{graphicx}
\usepackage{caption}
\usepackage{float}
\usepackage{tabularx}
\usepackage{array}
\usepackage{multirow}
\usepackage{makecell}
\newcolumntype{P}[1]{>{\centering\arraybackslash}p{#1\textwidth}}
\newcolumntype{L}[1]{>{\arraybackslash}p{#1\textwidth}}
\usepackage[table]{xcolor}

\usepackage{lineno}
\RequirePackage{fix-cm}
\smartqed  
\usepackage{graphicx}
\newcounter{chapter}
\renewcommand\thesection{\thechapter.\arabic{section}}
\renewcommand\thesubsection{\thesection.\arabic{subsection}}
\renewcommand\thesubsubsection{\thesubsection.\arabic{subsubsection}}
\renewcommand\thetable{\arabic{chapter}.\arabic{table}}
\renewcommand\thefigure{\arabic{chapter}.\arabic{figure}}
\begin{document}

\title{Review of the essential roles of SMCs in ATAA biomechanics}
%
%
\titlerunning{Review of the essential roles of smooth muscle cells in ascending thoracic aortic aneurysms biomechanics}        
\author{Claudie Petit \and S. Jamaleddin Mousavi \and St\'ephane Avril}
%
%
\institute{C. Petit$^{*}$ \and S. J. Mousavi \and S. Avril  \at
              Mines Saint-\'{E}tienne, Univ Lyon, Univ Jean Monnet, INSERM, U 1059 SAINBIOSE\\
              Centre CIS, F - 42023 Saint-\'{E}tienne France\\
              $^{*}$Corresponding author \email{claudie.petit@emse.fr}\\
              }
%
%
\maketitle
\begin{abstract}
Aortic Aneurysms are among the most critical cardiovascular diseases. The present study is focused on Ascending Thoracic Aortic Aneurysms (ATAA). The main causes of ATAA are commonly cardiac malformations like bicuspid aortic valve or genetic mutations. Research studies dedicated to ATAA tend more and more to invoke multifactorial effects. In the current review, we show that all these effects converge towards a single paradigm relying upon the crucial biomechanical role played by smooth muscle cells (SMCs) in controlling the distribution of mechanical stresses across the aortic wall. The chapter is organized as follows. In section \ref{passiverole}, we introduce the basics of arterial wall biomechanics and how the stresses are distributed across its different layers and among the main structural constituents: collagen, elastin, and SMCs. In section \ref{activerole}, we introduce the biomechanical active role of SMCs and its main regulators. We show how SMCs actively regulate the distribution of stresses across the aortic wall and among the main structural constituents. In section \ref{mechanosensing}, we review studies showing that SMCs tend to have a preferred homeostatic tension. We show that mechanosensing can be understood as a reaction to homeostasis unbalance of SMC tension. Through the use of layer-specific multiscale modeling of the arterial wall, it is revealed that the quantification of SMC homeostatic tension is crucial to predict numerically the initiation and development of ATAA.

\end{abstract}
%
%
\newpage
\section{Introduction}
Aortic aneurysms (AA) are among the most critical cardiovascular diseases \cite{bown2002,goldfinger2014}. Although their detection is difficult, prevention and monitoring of AA is essential as large AA present high risks of dissection or rupture, which are often fatal complications \cite{goldfinger2014,isselbacher2005}. Monitoring consists in measuring the aneurysm diameter using medical imaging methods such as echography or CT-Scan \cite{goldfinger2014,isselbacher2005}.
\\
The present study is focused on ascending thoracic AA (ATAA). The risk of rupture of ATAA is estimated clinically with the maximum aneurysm diameter, which consists in considering surgical repair for ATAA diameters larger than $5.5$ cm. Other factors such as growth rate, gender or smoking can be taken into account \cite{goldfinger2014,isselbacher2005,fillinger2007}. It is known that the criterion of maximum diameter relies on statistics of the global ATAA population. On an individual basis, many ruptures or dissections have been reported for aneurysms with diameters below the critical value \cite{fillinger2002}. Other criteria based on biomechanics were suggested \cite{fillinger2002}, but they still need to be validated clinically \cite{marquez2014,trabelsi2015,duprey2016}.
\\
The main causes of ATAA are summarized in Table~\ref{table1}. ATAAs are a very specific class of AA due to the particularity of the ascending thoracic aorta. First, it contains the highest density of elastic fibers of all the vasculature, which have to resist to the mechanical fatigue induced by the wearing combination of pulsed pressure and axial stretching repeated every cardiac beat. As elastic fibers cannot be repaired in mature tissue \cite{arribas2006}, the ascending aortic tissue is highly prone to mechanical damage \cite{isselbacher2005,marquez2014}. Second, a major role of the contractile function in smooth muscle cells (SMCs) is evident in the ascending aorta more than anywhere else as heterozygous mutations in the major structural proteins or kinases controlling contraction lead to the formation of aneurysms of the ascending thoracic aorta \cite{wolinsky1964}. Moreover, the outer curvature of the ascending thoracic aorta is constituted of a mix of Cardiac Neural Crest - and Second Heart Field -derived SMCs, distributed over the different Medial Lamellar Units (MLUs) (Fig. XX) \cite{sawada2017}. This may be correlated with the observation that dilatations are more often located on the outer curvature of the ascending thoracic aorta \cite{tremblay2010}. Third, the ascending thoracic aorta experiences very complex flow profiles, with significant alterations (vortex, jet flow, eccentricity, peaks of wall shear stress) in case of bicuspid aortic valves \cite{choudhury2009,pasta2013,michel2018} or aortic stenosis \cite{barker2012,schnell2016,michel2018}. It was shown that these complex hemodynamics patterns have major interactions with the aortic wall and correlate with local inflammatory effects or variations of oxidative stress in the aortic tissue \cite{miller2002,dua2010,michel2018}.

Research studies dedicated to ATAA have always invoked one of the three previous particularities of the ascending thoracic aorta to account for the intrinsic mechanism leading to the development of an ATAA, even if recent studies tend more and more to invoke multifactorial effects. In this review, we show that all these effects converge towards a single paradigm relying upon the crucial biomechanical role of SMCs in controlling the distribution of mechanical stresses across the different components of the aortic wall.  The chapter is organized as follows. In section \ref{passiverole}, we introduce the basics of arterial wall biomechanics and how the stresses are distributed across its different layers. In section \ref{activerole}, we introduce the biomechanical active role of SMCs and its main regulators and show how this can control the distribution of stresses across the aortic wall. In section \ref{mechanosensing}, we review the different pathways of SMCs mechanotransduction and their mechanisms at the cellular and tissue level in the aortic wall. Finally, we review studies showing that SMCs tend to have a preferred homeostatic tension. We show that mechanosensing can be understood as the reaction to homeostasis unbalance of SMC tension. The review reveals though that the quantification of the SMC homeostatic tension in the ascending thoracic aorta is still an open question and the chapterF closes with possible directions for research in measuring this tension at the tissue level and at the cellular level.

\section{Basics of aortic wall mechanics and passive biomechanical role of SMCs}\label{passiverole}

\subsection{Composition of arteries}\label{multiscalepb}
\subsubsection{The extracellular matrix (ECM)}\label{ECM}
The ECM of the aortic tissue is made of two main fibrous proteins participating to                                 the passive response: collagen and  elastin, which are responsible for $60\%$ in dry weight of the entire wall \cite{humphrey2015}. There are several types of collagen, but types I, III and V are primarily found in the media layer (see section below describing the layers of the aorta), where the SMCs are located, representing about $35\%$ of global aortic wall in dry weight \cite{humphrey2002,humphrey2015}.  Collagen fibers are not extensible and ensure the mechanical resistance of the tissue in case of overloading \cite{humphrey2015, wagenseil2009,wolinsky1964}. If collagen fibers can be produced over the lifespan, elastic fibers are actively synthesized in early development, and there is a loss of efficiency for the ones created during adulthood \cite{arribas2006}. Elastin has 40 years estimated half-life. It should last for an entire life in optimal conditions, but some pathological states or natural aging will necessarily affect it. Fibroblasts and SMCs can produce new ECM components but also matrix metalloproteases (MMPs) which degrade the current ECM. If the action of MMPs is not well regulated, the ECM may be remodeled, yielding a different mechanical behavior with possible ATAA development \cite{arribas2006, papke2015}. Likewise, the loss of elasticity may be related to an anomaly of elastic fibers. Elastic fibers are mainly composed of a core of amorphous elastin surrounded by microfibrills. The microfibrils comprise collagen VI and fibrillin \cite{kielty1992}, a polymer encoded by the \textit{fbn1} gene, whose mutation is involved in Marfan syndrome. The genetic mutations affecting the ECM in ATAAs are summarized in Table \ref{table1}.
\\
Another important constituent, although with lesser mass fractions, are Glycosamynoglycans (GaGs) which can contribute to the compressive stiffness of the aortic tissue. As they represent about $3$ to $5\%$ of the total wall by dry weight \cite{humphrey2015}, they do not participate markedly to the passive response except in specific cases, like for atherosclerosis, where GaGs are piled up during lesion development and increase therefore the wall stiffness. GaGs refer to different types of non-sulfated (hyaluronic acid) and sulfated (keratan sulfate, dermatan sulfate, and heparan sulfate) polysaccharides \cite{akgul2014}. The ECM contains also some glycoproteins that bind to cell membrane receptors, the integrins, and allow for cellular adhesion. Among these binding proteins, the fibronectin can also bind to collagen and heparan sulfate, and laminin is a major component of basal lamina which influences cell responses. 
\subsubsection{A multilayered wall structure}\label{layers}
The aortic wall is divided into three main layers surrounding the lumen where the blood flow circulates (Fig. XX). Each layer has its function and proper mechanical properties \cite{finet1999, humphrey2002,marquez2014,pasta2013,rapp1997,wagenseil2009}. The adventitia, which is the most external layer, contains fibroblasts and it is particularly collagen-rich, according to its protective role for the entire wall against high stress. The internal layer, called the intima, is directly into contact with the blood flow and constitutes a selective barrier of endothelial cells for preventing the wall from blood products infiltration and delivering oxygen and nutrients from blood to the internal wall. The inner medial layer is separated from adventitia and intima by two elastic laminae, and represents about $\frac{2}{3}$ of the whole thickness of the wall. All these layers have a passive mechanical response to the loading induced by the blood flow, but only the media can also act actively, due to the presence of contractile SMCs. The media is structured into several MLUs (Fig. XX) \cite{humphrey2015,rapp1997} where a layer of SMCs is tight between two thin elastin sheets, through a complex network of interlamellar elastin connections \cite{wagenseil2009}. The SMCs are oriented in the direction of the ECM fibers in order to better transmit the forces to each other and to successive MLUs. The number of MLUs varies according to the diameter of the artery \cite{rapp1997} and the size of the organism : $6-8$ for mice and $40-70$ for human body \cite{humphrey2015}.

\subsection{Basics of aortic biomechanics}\label{wallparameters}

It is commonly assumed that only the adventitia and the media are involved in the mechanical response of the entire wall, neglecting the mechanical role of the intima. This assumption is not valid in the case of pathologies resulting in a thickening of the intima like atherosclerosis.
\\
The aorta is submitted to four types of mechanical stresses (Fig. XX). The two main components are the axial one, $\sigma_z$, and the circumferential one, $\sigma_{\theta}$. The two other components are, namely $\sigma_r$ (radial stress) and $\tau_w$ (wall shear stress). The wall shear stress results from the friction of the blood onto the wall. 
The circumferential stress is related to the distension of the aorta with the variation of the blood pressure. It can reach about $150$ kPa under normal conditions \cite{humphrey2015}.  It can be approximated by the Laplace law according to:
\begin{equation}\label{sigtheta1}
\begin{aligned}
{\sigma_\theta} = \frac{P \cdot r}{t}
\end{aligned}
\end{equation}
where $\textit{P}$ is the blood pressure, $\textit{r}$ the internal aortic radius and $\textit{t}$ the thickness of the wall. If the number of MLUs varies according to the arterial diameter and across species \cite{wolinsky1967}, the average tension per MLU was shown to remain constant at $T=2$ N/m \cite{humphrey2015}, and its average circumferential stress can be determined by:
\begin{equation}\label{sigtheta2}
\begin{aligned}
{\sigma_\theta} = \frac{T}{t_{MLU}}
\end{aligned}
\end{equation}
As the mean thickness of a MLU is about ${t_{MLU}} \simeq {15}$ $\mu$m, it was estimated that the average normal circumferential stress across the aorta is ${\sigma_\theta} = 133$ kPa \cite{humphrey2015}.%
\subsection{Passive mechanics of the aortic tissue}\label{tissuemechanics}
The passive behavior refers to the behavior of the aortic wall in absence of vascular tone. It is mainly due to ECM components, namely elastin and collagen fibers. If the elastin is responsible for the wall elasticity, the collagen fibers are progressively tightened from their initial wavy configuration while the wall stress is increasing, and they tend to protect the other components from overstress \cite{humphrey2015,wagenseil2009,wolinsky1964}.
\\
Given that the tissue contains about $70$ to $80\%$ of water, it is often assumed as incompressible. As a heterogeneous composite material comprising a fluid part (\textit{i.e.}, water) and a solid part (\textit{i.e.}, ECM and cells) \cite{humphrey2002,pasta2013}, divided into several layers with different mechanical properties (see section \ref{layers}), the aortic wall has a complex anisotropic mechanical behavior. To predict the rupture risk of ATAAs \cite{humphrey2002,pasta2013,wagenseil2009}, the passive mechanical behavior of the ECM is relevant. Numerous \textit{in vitro} tests using the bulge inflation device \cite{brunon2011,duprey2016,kim2012,romo2014,trabelsi2015,cavinato2017} confirmed that elastin in the media is the weak element of the wall towards rupture.

\subsection{Multilayer model of stress distribution across the wall}
Single-layered homogenized models of arterial wall mechanics have provided important visions of arterial function. For example, Bellini \textit{et al.} \cite{bellini2014} proposed a bi-layer model with different material properties for the media and adventitia layers. They split the passive contributions of elastin, SMC and collagen fibers (modeled with four different families). Eventually, the strain-energy function (SEF) at every position may be written as \cite{bellini2014,humphrey2008}: 
\begin{linenomath*}
\begin{equation}\label{SEF}
W = \rho^{\text e} {W}^{\text e}({I}_1^{\text e}) + 
\sum_{i=1}^{n} \rho^{{\text c}_i} {W}^{{\text c}_i}({I}_4^{{\text c}_i}) + \rho^{\text m} {W}^{\text m}({I}_4^{\text m})
\end{equation}
\end{linenomath*}
where superscripts e, c$_i$ and m represent respectively the elastin fiber constituent, the constituent made of each of the $n$ possible collagen fiber families and the SMC constituent, all these constituents making the mixture. In Eq.~\ref{SEF}, $\rho^j$ refers to mass fraction, ${W}^j$ stands for stored elastic energy of each constituent, depending on the first ($I_1^j$) and fourth ($I_4^j$) invariants of the related constituents of the mixture ($j \in \lbrace {\text e}, {\text c}_i, {\text m} \rbrace$). 
Let the mechanical behavior of the elastin constituent be described by a Neo-Hookean strain energy function as in \cite{bellini2014,cardamone2009,dorrington1977,holzapfel2000}
\begin{linenomath*}
\begin{equation}\label{SEF_EF}
W^{\text e}(I_1^{\text e}) = \frac{\mu^{\text e}}{2}(I_1^{\text e}-3)
\end{equation}
\end{linenomath*}
where $I_1^{\text e} = tr(\bf C^{\text e})$ and $\mu^{\text e}$ is a material parameter and has a stress-like dimension. $\bf C^{\text e} = {\bf F^{\text e}}^{\text T} \bf F^{\text e}$ denotes the right Cauchy-Green tensor where $\bf F^{\text e}={\bf F} {\bf G}^{\text e}_h$ is the deformation gradient of the elastin constituent. $\bf F$ is the corresponding deformation gradient of the arterial wall mixture and ${\bf G}^{\text e}_h$ is the deposition stretch of elastin with respect to the reference configuration  \cite{bellini2014,cardamone2009}. Therefore, using the concept of constrained mixture theory (CMT) it is assumed that all constituents in the mixture deform together in the stressed configuration while each constituent has a different ``total'' deformation gradient based on its own deposition stretch.

The SEF of passive SMC and collagen contributions is described using an exponential expression such as \cite{bellini2014,cardamone2009,riveros2013,rodriguez2008}:
\begin{linenomath*}
\begin{equation}\label{SEF_CF_SM}
W^k(I_4^k) = \frac{D_1^k}{{4D_2^ k}} \left[ {\text {exp}} \left( {D_2^ k} (I_4^k -1)^2 \right) - 1 \right] 
\end{equation}
\end{linenomath*}
where $k \in \lbrace {\text c}_i, {\text m} \rbrace$. $D_1^{\text k}$ and $D_2^{\text k}$ are stress-like and dimensionless material parameters, respectively, and can take different values when fibers are under compression or tension \cite{Bersi2016}. $I_4^k = {G^k_h}^2 {\bf C} : {\bf M}^k \otimes {\bf M}^k$
where $G^k_h$, $k \in \lbrace {\text c}_i, {\text m} \rbrace$, is the specific deposition stretch of each collagen fiber family or SMCs, with respect to the reference configuration. ${\bf M}^k$, $k \in \lbrace {\text c}_i, {\text m} \rbrace$, denotes a unit vector along the dominant orientation of anisotropy in the reference configuration of the constituent made of the $i$th family of collagen fibers or of SMCs. For SMCs, ${\bf M}^m$ coincides with the circumferential direction of the vessel in the reference configuration while for the $i$th family of collagen fibers ${\bf M}^{{\text c}_i} = [0 \,\,\, {\text{sin}}\alpha^i \,\, {\text{cos}}\alpha^i]$, where $\alpha^i$ is the angle of the $i$th family of collagen fibers with respect to the axial direction. ${\bf C} = {\bf F}^{\text T} {\bf F}$ is the right Cauchy-Green stretch tensor of the arterial wall mixture \cite{bellini2014,cardamone2009}.

This model can capture the stress ``sensed'' by medial SMCs and adventitial fibroblasts. The model shows interestingly that the stresses spit unevenly between the media and the adventitia (Fig. XX). For physiological pressures, the stress is signifcantly larger in the media but when the pressure increases, the stress increases faster in the adventitia. As this chapter is dedicated to SMCs, the model permitted to estimate that stresses taken by SMCs remain less than a modest $40$ kPa for normal physiological pressures \cite{bellini2014}. 
\section{Active biomechanical behavior}\label{activerole}
On top of its passive mechanical behavior, the aortic tissue exhibits an active component thanks to the tonic contraction of SMCs, permitting fast adaptation to sudden pressure variation during cardiac cycle.
\subsection{Smooth Muscle Cells (SMC)}\label{SMC}
\subsubsection{SMCs structure}
SMCs have an elongated, fiber-like shape. Their length is about $50-100$ $\mu$m and their mean diameter is $3$ $\mu$m, reaching $5$ $\mu$m around the nucleus \cite{fujiwara1992,humphrey2002,mecham1995,miller1986}. SMCs have an axial polarity. Their longest axis tends to align with the direction of the principal stress applied to the ECM (Fig. XX). Each MLU in the aorta contains a layer of SMCs that are connected to the elastic laminae thanks to microfibrills \cite{humphrey2015,wagenseil2009}. They are circumferentially arranged throughout the media \cite{fujiwara1992,humphrey2002,wagenseil2009} and are particularly sensitive to $\sigma_z$ and $\sigma_\theta$ components of the wall stress \cite{li2007}. This specific structure may also explain the fact that the media has been revealed stronger circumferentially than longitudinally \cite{pasta2013} and that the forces produced by the SMCs are maximized in this direction \cite{fujiwara1992}. This ability of endothelial cells and SMCs to align along the direction of the applied stress has been confirmed by a number of \textit{in vitro} studies \cite{bao2003,dartsch1986,kanda1994}.
\\
The arrangement of  SMCs in the media used to be controversial \cite{oconnell2008}. The most recent studies (1980\textquotesingle s \cite{clark1985,wolinsky1967}, 1990\textquotesingle s \cite{gaballa1998}, 2000 \cite{dingemans2000}) describe SMC orientation as circumferential whereas a helical and oblique disposition was reported earlier (1960\textquotesingle s \cite{keech1960}, 1970\textquotesingle s \cite{osborne-pellegrin1978}). Fujiwara \& Uehara showed an oblique orientation in 1992 \cite{fujiwara1992}. Likewise, data are controversial about alignment parallelly to the vessel surface: Clark \& Glagov \cite{clark1985} agree with this statement unlike Fujiwara \& Uehara \cite{fujiwara1992}. Furthermore, some authors mention a change of SMC orientation in each subsequent MLU, creating a herringbone-like layout \cite{davis1993,dingemans2000}. Humphrey suggested the SMCs are oriented helically, closer to a circumferential direction \cite{humphrey2002}, but O\textquotesingle Connell suggested the SMCs may also be slightly radially tilted \cite{oconnell2008}.
\\
A recent study pointed out the importance of the helical disposition, suggesting that SMCs are oriented according to two intermingled helices \cite{tonar2015}. This disposition was assumed in several tissue models \cite{holzapfel2002,holzapfel2001}. Moreover, a tissue model for coronaries taking into account the orientation of SMCs suggested they contribute both to circumferential and axial stresses and tend to reorient towards the circumferential direction when blood pressure is increased \cite{chen2013}. Other studies \cite{clark1985,hayashi2012} suggested that the almost circumferential orientation is only valid for inner MLUs of the ascending thoracic aorta because SMCs seem to orient more axially close to the adventitia. This pattern was also confirmed by Fujiwara \& Uehara \cite{fujiwara1992}.
\subsubsection{Principle of SMC contractility}\label{contractility}
The contractility of SMCs is their defining feature, thanks to a strongly contractile cytoskeleton. SMCs have a well-developed contractile apparatus organized in cross-linked actin bundles, regularly anchored into the membrane with dense bodies \cite{balas2002} (Fig. XX). This layout implies a bulbous morphological aspect during contraction \cite{fay1973}.
There may be two types of actin filaments in the same bundle. The thick filament serves as a support for myosin heads and permits sliding of thin filaments during contraction, defining a so-called ``contractile unit''. Thin filaments are made of Alpha Smooth Muscle Actin ($\alpha$-SMA), an actin isoform specialized in the increase of cellular traction forces \cite{chen2007,goffin2006,skalli1986}. This isoform is specific to certain cell types, namely SMCs and myofibroblasts \cite{tomasek2002}. The $\alpha$-SMA filaments are created from their rod-like form, synthesized and assembled when focal adhesions (FAs) undergo  high stresses \cite{chen2007,goffin2006}. Genetic mutations may affect the genes encoding the components of the contractile apparatus (Fig. XX) and lead to ATAAs (Table \ref{table1}).
\\
The main signaling pathways controlling SMC contraction are summarized in Fig. XX. More details about these pathways may be found in \cite{berridge2008,hill-eubanks2011,lacolley2012,milewicz2016,reusch1996,somlyo1994}. However, it is important to mention that SMCs contractility is controlled by the modulation of intracellular ionic calcium concentration $\lbrack Ca^{2+}\rbrack _i$. The SMC membrane has many invaginations called caveolae where extracellular $Ca^{2+}$ ions can enter the cell \cite{taggart2001}. The increase of  $\lbrack Ca^{2+}\rbrack _i$ triggers the contraction above a certain threshold, activating myosin chains \cite{hill-eubanks2011}. Some studies revealed  $\lbrack Ca^{2+}\rbrack _i$ is a reliable indicator of SMC contractility, because it increases from $100$ nM in the relaxed state to $600-800$ nM once fully contracted \cite{humphrey2002}. But Hill-Eubanks \textit{et al.} \cite{hill-eubanks2011} underlined later that a $400$ nM concentration is sufficient to cause a complete contraction.
\\
Calcium entries in the SMC after some stimuli resulting in membrane depolarization, widely studied \textit{in vitro} with electrical \cite{fay1973}, electrochemical \cite{karaki1984,murtada2016,schildmeyer2000}, chemical, or even mechanical stimulation \cite{balasubramanian2013,davis1992}. In fact, some of these studies suggested that SMCs undergo a progressive membrane depolarization as intraluminal pressure increases under normal conditions \cite{balasubramanian2013,davis1992,hill-eubanks2011}. But when the mechanical stimulation becomes higher than normal, some studies have also highlighted that the SMCs undergo more depolarization, resulting in an alteration of their reactivity \cite{demoudt2017}.
\\
The most common protocol used to control SMCs contraction \textit{in vitro} remains the addition of potassium ions $K^+$ from a $KCl$ solution with a $50$ to $80$ mM concentration, that depolarizes the membrane \cite{karaki1984,malmqvist1999,murtada2016,schildmeyer2000,demoudt2017}. The extracellular media must also contain a calcium concentration  $\lbrack Ca^{2+}\rbrack _e$ to cause the activation of myosin heads by calcium entry into the cell. This is the reason for adding $CaCl_2$ solution to the media \cite{malmqvist1999}, or immersing the cells in a physiological Krebs-Ringer solution \cite{murtada2016}. The latter has the advantage of keeping biological tissues alive.
Calcium entry is also regulated thanks to a cytosolic oscillator which allows periodic release of calcium from intracellular reservoirs (\textit{i.e.}, endoplasmic reticulum) \cite{berridge2008,berridge1988}. The frequency is highly dependent on external stimuli like neurotransmitters, hormones or growth factors. If its primary role is to induce a single cell contraction, the secondary role of the cytosolic oscillator is also responsible for membrane depolarization of neighboring cells, in order to synchronize the contraction of several SMCs \cite {berridge2008,brozovich2016}. The $Ca^{2+}$ signaling pathway was included in the mechanical cellular model of Murtada \textit{et al.} \cite{murtada2012} to model SMCs contractility.
\\
The Angiotensin II (Ang II) signaling pathway has been widely developed in mice models and its link with aneurysms is well explained by Malekzadeh \textit{et al.} \cite{malekzadeh2013}. It may lead to SMCs contraction and may be used as a vasoconstrictor agonist in mice models \cite{malekzadeh2013,michel2018} or for isolated cells by addition in a bath \cite{hong2015}. But the review of Michel \textit{et al.} suggests that angiotensin II may also damage the intima \cite{michel2018}. In this case, intimal degradation leads to the activation of other signaling pathways that have an influence on SMC tone. Moreover, some studies suggested very active biological role of the intima through the secretion of nitric oxide (NO) that is involved in a pathway controling cell relaxation \cite{hong2015,demoudt2017}. Accordingly, the intimal integrity seems to have a strong influence on cell contractile response.
\\
The filament overlap involved in SMC contraction creates a ``cross-bridge'' whose function has been early described by some subcellular models based on the sliding-filament theory \cite{hai1988,huxley1953}. The cross-bridges have been assumed to be based on contractility activation/deactivation cycles through phosphorylation of the contractile unit. That is what Dillon \textit{et al.} \cite{dillon1981} have called the ``latch state''  which was used later in association with the sliding filament theory to develop another subcellular model for the SMC contractile unit \cite{murtada2016}. Several other cellular models combine the proper active behavior of SMCs with the passive behavior of its ECM \cite{chen2013,tan2015,zulliger2004}. The SMC is protected from a too high lengthening thanks to the intermediate filaments (made of desmine), linking dense bodies together \cite{chen2007}.
\subsubsection{Intracellular connections}\label{cellconnections}
Each SMC is covered by a basal lamina, a thin ECM layer ($40-80$ nm \cite{humphrey2002}) comprising type IV collagen, glycoproteins and binding proteins ensuring cell adhesion: the fibronectin and the laminin. The basal lamina represents about $12$ to $50\%$ of the volume of SMCs. This lamina is open around the gap junctions to allow cell communication \cite{balas2002}. These junctions allow the cells to exchange electrochemical stimuli required to synchronize the contraction of the whole MLU and to match with the successive MLUs \cite{challande2007}. SMCs are linked together thanks to thin collagen microfibrils permitting to transmit cell forces.
\\
Interactions between the media and the other layer (intima and adventitia) also have to be considered. The synchronization of the contraction is induced in the outermost MLU by their innervation thanks to the vasa vasorum present in the adventitia, and the nervous signal is transmitted to inner MLUs thanks to gap junctions. The vasa vasorum also  provides nutrients in the thickest arteries to complete the action of the intima for innermost SMCs \cite{humphrey2002}. Moreover, endothelial cells communicate with innermost SMCs (Fig. XX) secreting vasoactive agonists, neurotransmitters and GaGs, notably the Heparan Sulfate, which seems to influence the quiescence of the SMCs \cite{rubbia1989,tran-lundmark2015}. Further information about this topic can be found in \cite{lilly2014}.
\subsection{Multiscale mechanics of SMC contraction}\label{multiscalemeca}
\subsubsection{Subcellular behavior}\label{stressfiber}

Many experiments were developed to characterize SMCs traction forces thanks to $\lbrack Ca^{2+}\rbrack _i$ measurements \cite{dillon1981} or Traction Force Microscopy (TFM) techniques, from common substrate deformation methods \cite{balasubramanian2013,goffin2006,tolic2002}  to uncommon specific microdevices \cite{hall2017,tan2003,zhang2014}. SMC stiffness is closely linked to their contractile state \cite{hong2015,smith2005}. The reported values depend strongly on the measurement method. Common magnetic twisting cytometry gives a range of $\lbrack 10^0 - 10^2\rbrack$ Pa against $\lbrack 10^3 - 10^5\rbrack$ Pa for Atomic Force Microscopy (AFM) \cite{laurent2002}. Published stiffness and traction force values for SMCs are reported in Table \ref{table2} .

If the AFM was mainly used on the ECM of aneurysm samples \cite{chen2015,lindeman2010}, only Crosas-Molist \textit{et al.} \cite{crosas-molist2015} characterized aortic SMCs using AFM and showed an increase of their stiffness in the Marfan syndrome (from $3$ kPa for healthy tissue to $7$ kPa for pathological one). Interestingly, another team focused on rat vascular SMCs (without aneurysm) and tested them by AFM indentation with a functionalized tip to measure the adhesion forces to type I collagen. This work suggested that contracted (with Ang II) or relaxed SMCs regulate their focal adhesions \cite{hong2015}. 
\subsubsection{(Sub)cellular models for the SMC}\label{subcellmodel}
A common mechanical model of the SMCs and their contractile apparatus is the sliding filament theory. The original sliding filament theory permitted to model the $\alpha$-SMA filaments sliding on myosin heads. It was initially published by Huxley \& Huxley in 1953 \cite{huxley1953}. The filament overlap (\textit{i.e.}, thin filaments linked to thick filaments by myosin heads, see section \ref{contractility}) creates a ``cross-bridge'' modeling the contractile unit of a single SMC.
\\
Another important study was those of Dillon \textit{et al.} \cite{dillon1981} where the latch state was introduced to describe the activation/blockage cycles of the contractile unit through phosphorylation process. In low phosphorylation states, the active force can be maintained by the SMCs \cite{hai1988}. Dillon \textit{et al.} \cite{dillon1981} also highlighted that SMCs generate a maximal force when stretched at an optimal length. Gradually, further models took into account the orientation of the SMCs in the media \cite{chen2013} and the interaction between the cell and its ECM  \cite{rachev1999,tan2015,zulliger2004}. Only Murtada\textquotesingle s model  \cite{murtada2012} has integrated the regulation of $\lbrack Ca^{2+}\rbrack _i$ controlling SMC contraction (see section \ref{contractility}).
\subsection{Effect of SMC contraction on the distribution of stresses across the aortic wall}\label{hfgetfetcfsrezy}
The effects of SMCs contraction on the stress distribution across the wall were investigated in several studies, which the opening angle experiment. This permitted to assess the intramural stress induced by SMC contractility \cite{matsumoto1996,rachev1999}. Indeed the opening angle experiment reveals residual stresses, which can be related to passive ECM mechanics and to SMCs active contraction \cite{matsumoto1996,rachev1999}. 

It was shown that at physiological pressure, the pure passive response of the wall does not ensure uniform stress distributions, suggesting an essential role of the basal tone of SMCs to maintain a uniform stress distribution \cite{matsumoto1996,rachev1999}. But under the effects of a vasoactive agonist, SMCs contract through myogenic response and can provoke a rise of intraluminal pressure up to $200$ mmHg. As they change the intraluminal pressure, SMCs may also induce nonuniform stress distributions across the wall \cite{matsumoto1996}. 

In summary, SMCs are very sensitive to mechanical stimuli. They tend to keep the intramural circumferential stress as uniform as possible for physiological variations of the blood pressure, but the stress becomes non uniform for higher pressures \cite{matsumoto1996}. They adapt their myogenic response, ranging from $50$ kPa for the basal tone under normal physiological conditions to $100$ kPa for maximal SMC contraction \cite{rachev1999}.

\section{Mechanosensing and mechanotransduction}\label{mechanosensing}
Given its highly sensitive cytoskeleton and focal adhesions (Fig. XX), SMCs represent real sensors of the local mechano-chemical state of the ECM. Many experimental models have permitted to investigate this mechanosensing role and how it is involved in ATAAs and dissections  \cite{michel2018}. One of the main response to stimuli mechanosensing is mechanotransduction \cite{park2011,parker2002,saez2007,ahmed2015,livne2016}, which is the process of transducing wall stress stimuli into tissue remodeling \cite{humphrey2002,li2007,thyberg1995,humphrey2015,wang2007}.
\subsection{Mechanosensing}\label{cellsensor}
Many recent studies have highlighted the effects of the environment on SMCs response, in terms of protein synthesis, proliferation, migration, differentiation or apoptosis, thanks to its mechanosensitive architecture \cite{ahmed2015,livne2016,park2011,parker2002,petit2017,saez2007}. 
Mechanosensing relies on links between the ECM, focal adhesions and the cytoskeleton. The microfibrils provide an adhesive support to the SMCs through collagen VI \cite{kielty1992}. Because of this, when the microfibrils are damaged, SMCs sense an increase in stiffness and are no more able to transmit forces to each other through elastic fiber. According to several studies, the elastin acts for the maturation of the contractile apparatus of the SMCs and may encourage their quiescent phenotype \cite{arribas2006,karnik2003,stadler1989}.  In the ECM, two proteins are mainly involved in mechanosensing: fibronectin and laminin. Fibronectin is known for being mainly present in the ECM of blood vessels during early development and seems to encourage SMCs proliferation and migration in order to build the tissue \cite{reusch1996,thyberg1990}. On the contrary, the laminin may be required later for SMCs maturation towards a contractile phenotype \cite{tran2006}. 

Since SMCs are dynamic systems, their cytoskeleton remains in constant evolution during cellular processes. This specific structure allows the cell for shape maintenance and generation of traction forces required notably during migration (Fig. XX). The cytoskeleton of SMCs is particularly rich in contractile $\alpha$-SMA thin filaments (see section \ref{SMC}) that are used to enhance traction forces required for the cell function. SMC contraction involves a quick remodeling of its cytoskeleton in order to recruit contractile thin filaments in the direction of applied forces \cite{chen2007,goffin2006} and to follow its change of shape while renewing non-contractile cortical structures \cite{hong2015}.
\\
In summary, the SMC may be considered as a powerful sensor of the mechanical state across the aortic wall. The high sensitivity of SMCs led many research teams to point out their implication in arterial disease, including aortic aneurysms \cite{brownstein2017,guo2007,humphrey2002,karnik2003,li2007,mecham1995,thyberg1990}.
\subsection{The key role of SMCs in ATAAs}\label{SMCkeyplayer}
The role of SMCs in the development of ATAAs is now well accepted \cite{fujiwara1992,wagenseil2009}. 
\\
Several studies have already mentioned the change of SMC behavior in cardiovascular disease, and the consequences on the arterial wall. It was shown that hypertension is perceived by SMCs as permanent stimuli through the increase of wall stress, which induces collagen synthesis to reinforce the wall resulting in an increasing thickness \cite{michel2018,hayashi2009,humphrey2002,lacolley2012}.
\\
In atherosclerosis and restenosis, the growth of plaques between the media and the intima is due to SMCs proliferation and migration towards the intima, forming a neointima \cite{karnik2003,raines1993}. The neointimal SMCs are also able to gather lipids, increasing the stiffness and weakening the wall. Intimal integrity may also control the quiescence of SMCs thanks to Heparan Sulfate \cite{rubbia1989,tran-lundmark2015} or vasoactive agonists \cite{hong2015,humphrey2002} synthesis. Hence, the degradation of the endothelial cell layer leads to SMC proliferation and ECM synthesis until whole intimas repair \cite{thyberg1997}.
\\
All of these changes suggest that SMCs can switch to another phenotype, in order to repair the damaged tissue through migration towards the injured region, proliferation and ECM synthesis \cite{thyberg1997}. Under normal conditions, mature SMCs acquire a ``contractile'' $(C)$ phenotype from an immature ``synthetic'' $(S)$ one, which is mainly present in early development \cite{karnik2003,mecham1995,raines1993}. But SMCs demonstrate a high plasticity as they are not fully differentiated cells, and they can return to a $(S)$ phenotype in response to many stimuli. The phenotypic switching is due to a number of factors summarized Table \ref{table3} below.

The cytoplasm of $(S)$ SMCs has more developed synthetic organites like endoplasmic reticulum and Golgi apparatus, leading to hypertrophy \cite{humphrey2002}. The phenotypic switching does not radically change the cytoskeleton as microtubules remain intact, but the contractile apparatus (\textit{i.e.}, $\alpha$-SMA thin filaments) is clearly affected \cite{chen2007,karnik2003,rubbia1989,skalli1986,stadler1989,thyberg1997}. SMC contractility involves a reorganization of their contractile apparatus. In other words, high traction forces require high adhesion to the ECM, hence it is suggested that SMCs undergo a regulation of their focal adhesions \cite{anderson2004,chen2007,goffin2006,hong2015}, evolving towards super focal adhesions (suFAs) in the direction of the applied stress \cite{chen2007,goffin2006}. 
\\
Hyperplasia concerns the loss of SMCs' quiescence in favor of a proliferating and migrating behavior \cite{humphrey2002,mecham1995}. During hypertension, the increase in wall thickness has been shown to result in more from hypertrophy than hyperplasia \cite{lacolley2012,owens1981}, but the two phenomena are involved in several pathological states \cite{humphrey2002}. ATAAs also involve a reduction of the elastin/collagen ratio in the aortic wall, inducing stiffness increase and leading to phenotypic switching of SMCs \cite{mao2015}. But the whole thickness is not uniformly affected: Tremblay \textit{et al.} \cite{tremblay2010} have assessed SMC densities across ATAAs and deduced it was greater in the outer curvature. The reduced contractile behavior suggests more phenotypic switching in this area. 
\subsection{SMC mechanotransduction}\label{chain reaction}
As previously highlighted in several studies \cite{bellini2014,humphrey2015}, SMCs tend to remain in a specific mechanical state called homeostasis. It is considered as a reference value for the stress they undergo into the wall under normal physiological pressure. During any cardiac cycle, SMCs do not activate suddenly their contractile apparatus according to the short variations of blood pressure. In fact, they always remain partially contracted and tend to adapt gradually to any constant increase of the mean pressure (Fig. XX).

Facing a constant rise of wall stress, SMC response may be divided into two main categories according to time. In the short term, SMCs react in a progressive contraction until they reach maximal contraction, permitting to regulate the blood flow through arterial diameter control. But beyond a given stress threshold, collagen fibers from the adventitia are recruited to protect the cells and the medial elastic fibers from higher stress values \cite{humphrey2002,bellini2014}. In the long term, the remaining mechanical stimuli of SMCs lead to phenotypic switching or apoptosis. In this way, SMCs tend to coordinate the renewal of ECM, and particularly synthetizing collagen fibers to increase the wall resistance to high stress.
\subsection{Mechanical homeostasis in the aortic wall}\label{homeostasis}
Mechanical homeostatis means that SMCs try to regulate their contractile apparatus and their surrounding ECM to maintain a target wall stress corresponding to a certain mechanobiological equilibrium. The presence of a mechanobiological equilibrium was first proposed by the constant mean tension of a single MLU in a stressed aorta in spite of different species and aortic diameter \cite{humphrey2015,wolinsky1967}. Humphrey \cite{humphrey2015} estimated that the circumferential stress per MLU is about $\sigma _\theta = 133$ kPa. It is assumed that SMCs and fibroblasts tend to maintain a preferred mechanical state through homeostasis. Kolodney \textit{et al.} showed that cultured fibroblasts on unloaded gel substrates generate a steady tension of $3.2$ kPa \cite{kolodney1992}. Moreover, Humphrey \cite{humphrey2015} suggested that homeostasis expression is similar throughout scales, from organ level (vessel mechanoadaptation), tissue level (ECM prestressing and synthesis/degradation), cellular level (traction forces applied onto the ECM,  see section \ref{stressfiber}), subcellular level (focal adhesions and actin/myosin bundles) and even molecular level ($\lbrack Ca^{2+}\rbrack _i$). These findings suggest that the cell is able to adapt its proper stress state through the regulation of $\lbrack Ca^{2+}\rbrack _i$, cytoskeleton and focal adhesions turnover, and by controlling its surrounding ECM as well. Experimental studies of Matsumo \textit{et al.} \cite{matsumoto1996} showed a change of the intramural strain distribution in response to SMCs contraction (and relaxation) on radially cut aortic rings and confirm that SMCs actively adapt their contractile state to keep the intramural stress uniform. In summary, SMCs can both work actively (through contraction/relaxation) and passively by deposition and organization of the ECM \cite{humphrey2015}.
\\
That is why SMCs may undergo a phenotypic switching towards a synthetic one under several stimuli (see section \ref{SMCkeyplayer}). Through phenotypic switching, SMCs tend to remodel their ECM to go back to a preferred state and facing the variations of their environment. Humphrey has well described this equilibrium state saying: \textit{``When a homeostatic condition of the blood vessel is disturbed the rate of tissue growth is proportional to the increased stress''} \cite{humphrey2002}. But SMCs lose their contractility in return and may irrevocably affect the wall vasoactivity in which they may have a key role \cite{humphrey2015,kuang2012,schildmeyer2000}. In fact, Humphrey suggests in his review that fully contractile SMCs can react mainly to circumferential wall stress ($150$ kPa in physiological conditions) with a $100$ kPa equivalent traction forces exerted on their ECMs while synthetic SMCs may only apply $5-10$ kPa \cite{humphrey2015}.
\subsection{Consequences for aortic tissue}\label{consequences}
Reduction or loss of SMC contractility alters the stress distribution across the aortic wall \cite{hong2015,owens2004,skalli1986,stadler1989}. In reaction, the development of synthesis abilities ensures recovery processes by ECM remodeling. SMCs keep a key role in the aortic wall remodeling. In ATAAs, they tend to adapt their response through complex signaling pathways. An important one is Rho kinase (ROCK), mainly involved in cytoskeleton turnover for the control of cell shape and movement during migration \cite{kuang2012}. The Rho kinase seems to influence the formation of  $\alpha$-SMA thin filaments and the regulation of FAs that are involved in SMCs contractility and anchoring to the ECM \cite{anderson2004,chen2007,somlyo2003,goffin2006}.  Moreover, the oxidative stress induced by ATAAs enhances the inflammatory response of SMCs, increasing MMP synthesis and further disruption of elastin fibers \cite{miller2002}. 

Remodeling was shown to be uneven in human \cite{choudhury2009} and porcine \cite{tremblay2010} aortic tissues. Authors highlighted that the outer curvature of ATAAs is more affected. Remodeling implies phenotypic switching towards a synthetic phenotype able to synthesize both ECM compounds (\textit{i.e.}, collagen and glycoproteins) and MMPs to degrade the ``dysfunctional'' ECM, leading to ECM wear \cite{humphrey2002,owens2004,raines1993,rubbia1989}. Likewise, elastin degradation results in a permanent decrease of the elastin/collagen ratio since elastic fibers cannot be regenerated in adulthood \cite{arribas2006}. On top of the induced stiffening, the ability of SMCs to restore an healthy state is altered as well as it was shown that elastin is also important for activating actin polymerization \cite{wang2007}. Moreover,  SMCs undergo a general apoptosis to reduce their number when they sense an unappropriate chemo-mechanical state, inducing further reduction of elasticity and mechanical resistance through a vicious circle loop \cite{miller2002,li2007,mao2015,riches2013}.
\subsection{Towards an adaptation of SMCs in ATAAs?}\label{adaptation}
Any disruption of the mechanical or chemical homeostasis is interpreted by the SMCs as a distress signal, and several recovery processes can be activated in reaction, but the regulation loop is similar to a vicious circle because of the complexity to return naturally to equilibrium (Fig. XX).
Interestingly, in hypertension, the increase of wall stress results in an increase in the arterial diameter \cite{hayashi2009,humphrey2002}. Conversely, a decrease in mean wall stress leads to an atrophy \cite{humphrey2002}. Since the $(S)$ SMCs can recover their $(C)$ phenotype once the tissue returns to its original homeostatic stress, the phenotypic switching seems to be a reversible process \cite{owens2004,stadler1989,thyberg1995}. These observations suggest a two-way mechanoadaptive process.\\
But once affected by ATAAs, remodeled aortic ECM is known not to reach complete recovery, particularly because disrupted elastic fibers can not be rebuilt in adulthood \cite{arribas2006}. Aortic tissue would, therefore, evolve more or less quickly according to some factors that may slow it down. As the review study of Michel \textit{et al.} \cite{michel2018} has already pointed it out, ATAAs may result in some epigenetic modifications that have an influence on the cellular response. It could be defined as the acquisition of new constant and heritable traits without requiring any change in the DNA sequence, which results for instance in gene modulation. The suggested theory explains that SMCs reprogramming is likely to induce a progressive dilatation of the aorta without dissection, whereas no reprogramming SMCs promote acute rupture of the wall \cite{michel2018}. Finally, it is well accepted that SMCs play a major role in controlling the wall evolution after aortic injury, either toward partial recovery of initial mechanical properties or fatal rupture through dissection. However, quantification of levels of SMC contractility that result in one type of evolution or another is still an open issue. There is still a pressing need to characterize the basal tone of SMCs in healthy aortas ATAAs at the cellular scale.
\\

\section{Summary and future directions}
Mechanobiology and physiopathology of the aorta have received much attention so far but there is still a pressing need to characterize the roles of SMCs at the cellular scale. Despite the difficulties of characterizing cells having complex dynamics and fragility, techniques such as Atomic Force Microscopy or Traction Force Microscopy could permit important progress about SMC nanomechanics and provide relevant information on how SMC biomechanics is related to the irreversible alteration of stress distribution in ATAAs. This will also imply the development of new biomechanical models of the aortic wall taking into account the contractility of SMCs. 
\section{Acknowledgments}
The authors are grateful to the \textit{European Research Council} for grant \textit{ERC-2014-CoG BIOLOCHANICS}.
\section{Conflict of interest}
There is no conflict of interest.
\medskip
\bibliographystyle{elsarticle-num}
\bibliography{BiblioReview}

\newpage
\listoftables
\captionsetup[table]{labelfont={bf},labelformat={default},labelsep=period,name={Table}}
\newpage
\begin{table}[H]
\begin{center}
\caption{Main causes of ATAAs affecting both the ECM and the SMCs}\label{table1}
\begin{tabular}{|p{3.5cm}p{6cm}p{2cm}|}
\hline
\textbf {Causes of ATAAs} & \textbf {Effects} & \textbf {Ref.} \\ 
\hline
\multicolumn{3}{|c|}{\textit{Genetic mutations affecting the ECM}} \\
\hline {} & {} & {} \\
fnb1 (Marfan syndrome)   &  Microfibrils anomalies  => wrong force transmission and alteration of the mechanotransduction. & \cite{brownstein2017,isselbacher2005}  \\  {} & {} & {}
\\
Types I and III collagen   &  Anomalies of collagen fibers. & \cite{humphrey2015} \\ 

\hline
\multicolumn{3}{|c|}{\textit{Genetic mutations affecting the SMC}} \\
\hline   {} & {} & {}
\\
        
ACTA2 ($\alpha$ -SMA)  & Dysfunction of the contractile apparatus. This mutation represents about 12\% of ATAAs. & \cite{brownstein2017,guo2007}\\ {} & {} & {}
\\
MYH11 (Myosin light chain)  & Dysfunction of the contractile apparatus. & \cite{brownstein2017,kuang2012}\\ {} & {} & {}
\\
TGFB (TGF-$\beta$)  & Anomalies of TGFB receptors TGBFR1/2.
Wrong regulation of traction forces. & \cite{brownstein2017,chen2007,gillis2013} 
\\{} & {} & {}
\\
MYLK (Myosin light chain kinase)  & Alteration of myosin RLCs (regulatory light chains) phosphorylation, and thus force generation. & \cite{milewicz2016}\\{} & {} & {}
\\
PRKG1  & Kinase activation resulting in SMCs relaxation. & \cite{milewicz2016}\\{} & {} & {}
\\
MMP genes  & Alteration of myosin RLCs (regulatory light chains) phosphorylation, and thus force generation. & \cite{milewicz2016}\\

\hline
\multicolumn{3}{|c|}{\textit{Phenotypic switching : Contractile} (C) => \textit{Synthetic} (S) }\\
\hline {} & {} & {}
\\
Stiffening and weakening of the arterial wall.\newline \textbf{Pathologies} : atherosclerosis, arteriosclerosis, arteritis, aging. & The SMCs move on to synthetic phenotype (S) (hypertrophy), they lose their quiescence (hyperplasia), they order wall remodeling by the synthesis of MMPs (degradation) and ECM (renewal). Moreover, atheroma plaques contain many SMCs. & \cite{humphrey2002,isselbacher2005,mao2015,rubbia1989,tsamis2013}\\
\hline
Chronic overstress.\newline \textbf{Pathologies} : hypertension, dissection, ATAAs. & The (C) SMCs move on to (S) : remodeling, hypertrophy, hyperplasia. & \cite{arribas2006,owens1981}\\{} & {} & {}
 \\
\hline
Contact with blood flow.\newline \textbf{Pathologies} : intimal injury, porosity of the wall. & The (C) SMCs move on to (S), formation of a neointima containing SMCs and GAGs through hyperplasia. & \cite{thyberg1995}
\\
{} & Blood-borne components interacts with SMCs.
 & \cite{michel2018}
\\
\hline
Change in ECM chemical composition.\newline Laminin/fibronectin ratio. \newline Elastin/collagen ratio. & The (C) phenotype may be favored on laminin or Matrigel (collagen+laminin) in vitro. \newline A high elastin concentration may activate actin polymerization and thus the development of the contractile apparatus. & \cite{chen2015,reusch1996,thyberg1997,thyberg1990} \newline \cite{wang2007}\\
\hline
Cell culture in vitro.\newline  High passage. \newline
Substrate (physical properties). & Wrong development of the contractile apparatus => more (S) SMCs as cell passage increases. \newline
Necessity to use some stimuli like vasoactive agonists or suitable substrates. & \cite{murray1990,murtada2016,raines1993,stadler1989,thyberg1997}\\

\hline
\multicolumn{3}{|c|}{\textit{Partially identified causes} }\\
\hline {} & {} & {}
\\
Biochemical imbalance.\newline Signaling pathways involved in cell contraction.
& Angiotensin II,
Growth factors : TGF-$\beta$, PDGF.
& \cite{goldfinger2014,humphrey2015,li2007,milewicz2016,park2011,reusch1996,somlyo2003,tomasek2002}\\
{} & $Ca^{2+}$ ionic channel.
 & \cite{berridge2008,davis1992,hill-eubanks2011,malmqvist1999,murtada2016,schildmeyer2000,somlyo1994}\\
 \hline
Intercellular interactions.\newline Vasoactive agonists, neurotransmiters, hormones, ions, mechanical stimuli.
& Interaction with endothelial cells from the intima.
& \cite{hong2015,humphrey2002,rubbia1989,tran-lundmark2015,lilly2014,thyberg1997}\\
{} & Synchronization of several SMCs.
 & \cite{berridge2008,brozovich2016,humphrey2002}
 \\
 \hline
 Local changes in hemodynamics. \newline Bicuspid aortic valve, dissection, ATAAs.
& Disturbance of the mechanotransduction through endothelial cells and SMCs.
& \cite{isselbacher2005,michel2018,humphrey2002,owens2004}\\

\hline
 Embryonic origin of the SMCs. \newline Transition area between aortic root and arch : the media combines SMCs from different origin.
& Outermost SMCs are from second heart field and innermost ones from neural crest. This area is prone to ATAAs and dissections.
& \cite{sawada2017}\\

\hline
\end{tabular}
\end{center}
\end{table}
\newpage
\begin{table}[H]
\begin{center}
\caption{Stiffness and traction force values for in vitro SMCs}\label{table2}
\begin{tabular}{|P{0.3} P{0.05} L{0.5} L{0.1}|}
\hline
\textbf {SMCs mechanical properties} & \textbf {Value (Pa)} & \textbf {Description}  & \textbf {Ref.} \\
\hline 
\hline
\rowcolor{gray!60} \multicolumn{4}{|P{1.2}|}{\textbf{STIFFNESS}} \\
\hline
\hline
\textit{$10^{3}-10^{5}$ Pa} &  \multicolumn{2}{P{0.5}}{\textit{AFM technique}}   & \cite{laurent2002} \\
\hline
Viscoelastic properties   &  \multicolumn{2}{L{0.5}}{\textit{In response to a vasoactive agonist (serotonin)}} & \multirow{4}*{\cite{smith2005} (Airways SMCs)} \\
Storage modulus G\textquotesingle  &  $\longrightarrow$ & $150\%$ increase &  \\ 
Loss modulus G\textquotedbl   &  $\longrightarrow$ & $67\%$ increase & \\ 
Hysterisis & $\longrightarrow$ &  $28\%$ decrease after AFM stimulus : The cell elasticity prevails gradually more (\guillemotleft latch state\guillemotright) & \\
Elastic properties &  \multicolumn{3}{L{0.8}|}{\textit{Comparison between control an Marfan induced aneurysm tissue}}\\

\multirow{8}*{Young Modulus} & $3$k & \multirow{2}{6cm}{Increase in SMCs and ECM stiffness in the pathological case} & \multirow{3}*{\cite{crosas-molist2015}(aortic SMCs)} \\
 & $7$k & & \\
 & $\hookrightarrow$ & Increase in focal adhesions size & \\
 & \multicolumn{3}{L{0.8}|}{\textit{Comparison between control and stimulated tissue with Angiotensine II (vasoconstrictor)}} \\
 & $13.5$k & \multirow{2}{6cm}{Increase in SMCs stiffness after having their contraction induced (after 2 min)} & \multirow{6}{6cm}{\cite{hong2015} \* \cite{smith2005}} \\
 & $18.5$k & & \\
 & $22$k & \multirow{1}*{After 30 min (\* actin polymerisation dynamics)} &  \\
 & $\hookrightarrow$ & \multirow{3}{6cm}{Increase in focal adhesions size (Stronger adhesion to functionalized AFM tip with type I collagen)} & \multicolumn{1}{c|}{}
\\ \multicolumn{4}{|c|}{}
\\ \multicolumn{4}{|c|}{}
\\
\hline
$10^{0}-10^{2}$ Pa  &  \multicolumn{2}{P{0.5}}{\textit{Magnetic twisting cytometry}}   & \cite{laurent2002} \\
\hline
\multicolumn{3}{|L{0.8}}{\textit{Increase in SMCs stiffness with substrate rigidity}}   & \cite{hubmayr1996} \\
$12.6 \pm 1.6$ N/$m^{2}$ & $12.6$ & \guillemotleft Hard\guillemotright substrate : high density collagen & {} \\
$4.3 \pm 0.3$ N/$m^{2}$ & $4.3$ & \guillemotleft Soft\guillemotright substrate : low density collagen & {} \\
\multicolumn{3}{|L{0.8}}{\textit{Increase in SMC stiffness with contraction}}   & \cite{hubmayr1996} \\
{} & {} & More effect on \guillemotleft Soft\guillemotright substrate & {} \\
{} & {} & Increase linked to myosin heads activation and actin polymerisation  & \multirow{2}*{\cite{an2002} (Airways SMCs)} \\
$9.91 \pm 0.75$ N/$m^{2}$ & $9.9$ & Unstimulated &  \\
$14.27 \pm 0.85$ N/$m^{2}$ & $14.3$ & Vasoconstrictor agonist : serotonin & {} 
\\
\hline
\rowcolor{gray!60} \multicolumn{4}{|P{1.2}|}{\textbf{TRACTION FORCES}} \\
\hline
\hline
\multicolumn{4}{|L{1.2}|}{\textit{Traction forces measured according to the calcium concentration of the KCl bath}} 
 \\
 \hline
$2.9 \pm 0.4 \times 10^{5}$ N/$m^{2}$ & $290$k & $[Ca^{2+}]=1.6$ mM & \multirow{2}*{\cite{dillon1981} (Carotid SMCs)} \\
$3.9 \pm 0.2 \times 10^{5}$ N/$m^{2}$ & $390$k & $[Ca^{2+}]=25$ mM & \\
\hline
\multicolumn{4}{|L{1.2}|}{\textit{TFM on a PDMS micro needles array with a fibronectin coating, simulating a soft material : cellular stress applied by the entire cell}} \\
\hline
\multirow{3}{4cm}{SMCs applied stress, adhering to the pattern, unstimulated} & $4.6$k &  Inhibition of the myosin contractility and the actin polymerisation  & \multirow{6}*{\cite{tan2003}(Airway SMCs)} \\
\multirow{3}{4cm}{Increase in the applied stress per needle with cell spreading} & $10$k & Weakly spread ($440$ $\mu m^{2}$) & \\
 \multicolumn{4}{|L{1.2}|}{} \\
{} & $30$k & Strongly spread ($1520$ $\mu m^{2}$) & \\
 \multicolumn{4}{|L{1.2}|}{} \\
\hline
\multicolumn{4}{|L{1.2}|}{\textit{TFM on a gel substrate including fluorescent and magnetic microbeads with fibronectin coating: Traction forces measurement and mechanical stimulus}} 
\\
\hline
$1$ N/$m^{2}$ & $1$ & Unstimulated SMCs  & \multirow{2}*{\cite{balasubramanian2013} (Renal vSMCs)}  \\
$1.6$ N/$m^{2}$ & $1.6$ & Stimulated SMCs : 
$60\%$ increase & \\
\hline
\multicolumn{4}{|L{1.2}|}{\textit{Standard TFM on gel substrate with fluorescent microspheres : measurement of the deformation field after a chemical stimulus}} \\
\hline
\multirow{4}{4cm}{Increase in mean traction force (mean vector of the deformation field)} & $50$ & Unstimulated SMCs  & \multirow{4}*{\cite{tolic2002} (Airway SMCs)} \\
 \multicolumn{4}{|L{1.2}|}{} \\
{} & $100$ & Vasoconstrictor agonist : histamine & \\
 \multicolumn{4}{|L{1.2}|}{} \\
\hline
\multicolumn{4}{|L{1.2}|}{\textit{The contractile apparatus is made of non-contractile thick filaments, linked to classic focal adhesions (FA), and highly contractile α-SMA filaments, linked to super focal adhesions (suFA)}} \\
\hline
 \multirow{4}{4cm}{Increase in mean traction force (mean vector of the deformation field)} & $8.5$k & Stress produced by suFAs  & \multirow{4}*{\cite{goffin2006} (Myofibroblasts)} \\
 \multicolumn{4}{|L{1.2}|}{} \\
{} & $3.1$k & Stress produced by classic FAs & \\
\multicolumn{4}{|L{1.2}|}{} \\
\hline
\end{tabular}
\end{center}
\end{table}

\newpage

\begin{table}[H]
\begin{center}
\caption{SMCs phenotypic switching characteristics}\label{table3}
\begin{tabular}{p{3cm}p{5cm}p{2cm}ll}
\hline
\textbf {SMCs phenotypic switching}    & \textbf {Effects}    \\
\hline
\\
High plasticity   &  The SMCs are not entirely differentiated when they reach maturity through contractile (C) phenotype, and can move on to a synthetic (S) phenotype & \cite{humphrey2002,humphrey2015,mecham1995,owens2004,thyberg1997,thyberg1995,thyberg1990} \\
{} &
 (The (S) phenotype is mainly present in the aorta during early development)
 \\
\hline
\\       
ECM synthesis and degradation (through MMPs synthesis)  & The SMCs undergo an increase in volume (hypertrophy), with the development of their synthetic organites (Golgi apparatus and endoplasmic reticulum) & \cite{humphrey2002,owens2004,raines1993,rubbia1989}
 \\
\hline
\\
Loss of quiescence : hyperplasia  & (S) SMCs tend to proliferate and migrate & \cite{mecham1995}
\\
\hline
\\
Loss of contractility  & Stress produced into the wall :  & \cite{humphrey2015} 
\\
{} & (C) SMCs : $100$ kPa ; (S) SMCs : $5-10$ kPa & {}
\\
\hline
\\
Degradation of the contractile apparatus   & The cytoskeleton is not entirely remodeled (undamaged microtubules), but there are weaker actin and myosin concentrations (contractile fibers) in (S) SMCs & \cite{karnik2003,rubbia1989,stadler1989,thyberg1997}\\
\hline
\\
Modification of the basal side  & Regulation of the focal adhesions & \cite{hong2015,owens2004,stadler1989}
\\
\hline
\\
{} & (They grow according to the traction force direction, ensuring a strong adhesion to the ECM in response to high stress) & \cite{chen2007,goffin2006}
\\
Decrease in $\alpha$-SMA concentration  & Degradation of the thin filaments that are responsible for amplifying and regulating the cell traction forces & \cite{chen2007,skalli1986}
\\
\hline
\\
Reversible process & Once the tissue is repaired, the SMCs return to a contractile phenotype & \cite{owens2004,stadler1989,thyberg1995}
\\
\hline
\\
General apoptosis & Decrease of SMCs number and degradation of the ECM => loss of wall elasticity and resistance & \cite{li2007,mao2015,riches2013}
\\

\hline
\end{tabular}
\end{center}
\end{table}
\end{document}